\documentclass[prl,twocolumn,amsmath,preprintnumbers]{revtex4}
\usepackage{amsmath,amsfonts,bm}
\usepackage{graphicx}
\begin{document}

\title{Escorted Free Energy Simulations: Improving Convergence by Reducing Dissipation}

\author{Suriyanarayanan Vaikuntanathan$^1$ and Christopher Jarzynski$^{1,2}$}
\affiliation{$^1$Institute for Physical Science and Technology,University of Maryland, College Park, MD 20742\\ 
$^2$Department of Chemistry, University of Maryland, College Park, MD 20742}

\begin{abstract}

Nonequilibrium, ``fast switching'' estimates of equilibrium free energy differences, $\Delta F$, are often plagued by poor convergence due to dissipation.
We propose a method to improve these estimates by generating trajectories with reduced dissipation.
Introducing an artificial flow field that couples the system coordinates to the external parameter driving the simulation,
we derive an identity for $\Delta F$ in terms of the resulting trajectories.
When the flow field effectively escorts the system along a near-equilibrium path, the free energy estimate converges efficiently and accurately.
We illustrate our method on a model system, and discuss the general applicability of our approach.

\end{abstract}
\maketitle

The estimation of free energy differences is a challenging problem of central importance in computational thermodynamics.
The problem can be formulated as follows.
Given two equilibrium states of a system of interest, at the same temperature $\beta^{-1}$ but different values of an external parameter, $\lambda=A, B$, 
how do we estimate the corresponding free energy difference, $\Delta F = F_B - F_A$?
While the widely used {\it thermodynamic integration} and {\it perturbation} methods are based on equilibrium sampling,
recently there has been interest in the use of nonequilibrium simulations to estimate free energy differences~\cite{Frenkel,Chipot2007}.
In the most direct implementation of this approach, one repeatedly simulates a thermodynamic process during which the parameter $\lambda$ is ``switched'' at a finite rate from $A$ to $B$, with initial conditions sampled from equilibrium.
$\Delta F$ is then estimated using the identity~\cite{CJ:Equality}
\begin{equation}
\label{previousresult}
e^{-\beta\Delta F} = \left\langle e^{-\beta W}\right\rangle \approx
\frac{1}{N} \sum_{n=1}^N e^{-\beta W_n} .
\end{equation}
Here angular brackets denote an ensemble average over realizations of the process,
$W_n$ is the work performed on the system during the $n$'th of $N$ such simulations,
and the approximation becomes an equality as $N\rightarrow\infty$.

While Eq. (\ref{previousresult}) implies that we can determine $\Delta F$ using simulations of arbitrarily short duration (``fast switching''~\cite{Hendrix2001}), we pay a penalty in the form of poor convergence~\cite{Kofke,Gore,CJ:Rareevents}, as
the number of simulations needed to obtain a reliable free energy estimate using Eq. (\ref{previousresult}) increases rapidly with the {\it dissipated work},
\begin{equation}
\langle W_{\rm diss}\rangle \equiv \langle W\rangle - \Delta F \ge 0 ,
\end{equation}
that accompanies fast switching simulations.
This dissipation is a consequence of the second law of thermodynamics,
and reflects the {\it lag} that develops as the system pursues -- but is unable to keep pace with -- the equilibrium state corresponding to the continually changing value of the work parameter, $\lambda$~\cite{Lag1,Lag2,Hermans91}.
We can diminish the lag by running longer simulations, but this increases the computational cost per simulation.

In this Letter we introduce a general strategy for improving the efficiency of fast switching free energy estimates.
In our approach, the ``physical'' equations of motion ordinarily used during a simulation are modified by the addition of an artificial flow field, ${\bf u}({\bf z},\lambda)$, that {\it directly couples the evolution of the system coordinates ${\bf z}$ to variations in the work parameter, $\lambda$} (Eq. \ref{eq:modifiedEOM}).
Our central result, Eq. (\ref{modifiedequation}), is an identity for $\Delta F$ in terms of trajectories generated with the modified dynamics.
While this result is valid for an arbitrary, well-behaved flow field (reducing to Eq. (\ref{previousresult}) when ${\bf u}={\bf 0}$),
the method is particularly effective when this field is constructed so as to escort the system along a near-equilibrium path.
In particular, if ${\bf u}$ entirely eliminates the above-mentioned lag, then our method provides a perfect estimator of the free energy difference: $W = \Delta F$ for every simulation.

Consider a classical system described by a Hamiltonian $H({\bf z};\lambda)$, or $H_\lambda({\bf z})$, where ${\bf z}$ specifies a point in $d$-dimensional phase space (or configuration space).
At temperature $\beta^{-1}$, the equilibrium state of this system is described by the distribution
\begin{equation}
\label{eq:peq}
p^{\rm eq}({\bf z},\lambda) = \frac{1}{Z_\lambda} \exp[-\beta H({\bf z},\lambda)] ,
\end{equation}
with free energy $F_\lambda = -\beta^{-1} \ln Z_\lambda$.
We are interested in the difference $\Delta F = F_B - F_A$.

We suppose that we have a preferred set of equations of motion for simulating the evolution of the system, which we write in the generic form
\begin{equation}
\label{eq:physicalEOM}
\dot{\bf z} = \tilde {\bf v}({\bf z},\lambda),
\end{equation}
where $\dot{\bf z} = {\rm d}{\bf z}/{\rm d}t$, and $\tilde {\bf v}({\bf z},\lambda)$ typically contains both deterministic and stochastic terms.
Examples include Hamilton's equations, Langevin dynamics, and the Andersen and Nos\' e-Hoover thermostats~\cite{Frenkel}.
(While we treat time as a continuous variable in this Letter, our method can be generalized to include discrete-time Monte Carlo dynamics~\cite{discreteflow}.)
Eq. (\ref{eq:physicalEOM}) can be either stationary or explicitly time-dependent, according to whether we hold $\lambda$ fixed or vary it with time.
An ensemble of trajectories evolving under Eq. (\ref{eq:physicalEOM}) is described by a phase space density $f({\bf z},t)$ satisfying a Liouville-type equation,
\begin{equation}
\label{eq:liouville}
\frac{\partial f({\bf z},t)}{\partial t} ={\cal L}_\lambda \cdot f({\bf z},t) .
\end{equation}
As in Refs.~\cite{CJ:MasterEquation,Hummer}, we assume ${\cal L}_\lambda \cdot e^{-\beta H_\lambda} = 0$,
i.e.\ the equilibrium state is preserved when $\lambda$ is fixed.
We will use the term {\it physical dynamics} to refer to the evolution described by Eq. (\ref{eq:physicalEOM}) (at the single-trajectory level) or Eq. (\ref{eq:liouville}) (at the ensemble level), to emphasize that these dynamics are intended to model, to some degree of realism, the microscopic evolution of our system of interest.

Now suppose we modify Eq. (\ref{eq:physicalEOM}) by adding a term proportional to $\dot\lambda = {\rm d}\lambda/{\rm d}t$:
\begin{equation}
\label{eq:modifiedEOM}
\dot{\bf z} = \tilde {\bf v} + \dot\lambda \, {\bf u} ,
\end{equation}
where ${\bf u} = {\bf u}({\bf z},\lambda)$ is an arbitrary, continuous vector field on phase space.\footnote{
If ${\bf u}$ is not bounded, we must also impose a modest condition ``at infinity'', namely, $\lim_{z\rightarrow\infty} u e^{-\beta H} z^{d-1}= 0$.
}
With this additional, artificial term, every small increment of the work parameter, ${\rm d}\lambda$, induces a phase-space displacement, ${\rm d}{\bf z} = {\bf u}\,{\rm d}\lambda$.
Under these modified dynamics, the phase-space density satisfies
\begin{equation}
\label{Fevolution}
	\frac{\partial f}{\partial t}  = {\cal L}_\lambda f -\dot{\lambda} \nabla \cdot ({\bf u} f) \equiv {\cal L}^{\prime}_{\lambda,\dot{\lambda}} f ,
\end{equation}
where the continuity term $-\dot{\lambda} \nabla \cdot ({\bf u} f)$ accounts for the flow $\dot\lambda{\bf u}$.
We now derive our central result, Eq. (\ref{modifiedequation}), by generalizing the analysis of Hummer and Szabo~\cite {Hummer} to include the $\dot\lambda$-dependent terms in Eqs. (\ref{eq:modifiedEOM}) and (\ref{Fevolution}).

From Eq. (\ref{Fevolution}), we have
\begin{equation}
\label{Lannihilation}
	{\cal L}^{\prime}_{\lambda,\dot{\lambda}} \, e^{-\beta H}
	= \beta\dot{\lambda} \left[ {\bf u}\cdot(\nabla H) - \beta^{-1} (\nabla \cdot {\bf u} ) \right ]e^{-\beta H} .
\end{equation}
Now consider a specific protocol $\lambda_t$ for varying the work parameter from $\lambda_0=A$ to $\lambda_\tau=B$, and consider the {\it sink equation},
\begin{equation}
\label{sink}
	\frac{\partial g}{\partial t}={\cal L}^{\prime}_{\lambda,\dot{\lambda}} g -\beta \dot{\lambda} \frac{\partial\!\!\!/ H}{\partial\!\!\!/\lambda} g ,
\end{equation}
where we have introduced the compact notation
\begin{equation}
\label{eq:defSlash}
\frac{\partial\!\!\!/ H}{\partial\!\!\!/\lambda}({\bf z},\lambda)
\equiv
\frac{\partial H}{\partial \lambda}+{\bf u}\cdot\nabla H - \beta^{-1}  \nabla \cdot {\bf u} .
\end{equation}
Using Eq. (\ref{Lannihilation}) we verify by inspection that the function
\begin{equation}
\label {Boltzmann}
	g({\bf z},t)=\frac{1}{Z_A} \, e^{-\beta H({\bf z},\lambda_t)}
\end{equation}
is a solution of Eq. (\ref{sink}).
Independently, the Feynman-Kac theorem provides a path-integral solution of Eq. (\ref{sink}) \cite{Hummer,Hummer05,Ge08}.
Equating these two solutions, we get
\begin{equation}
\label{FK}
	\frac{1}{Z_A} \exp\left[-\beta H({\bf z},\lambda_t)\right]
	=\Bigl \langle  \delta({\bf z} - {\bf z}_t) \, \exp(-\beta w_t) \Bigr\rangle_{\bf u} ,
\end{equation}
where $w_t = \int_0^t {\rm d}t^\prime \dot\lambda\,(\partial\!\!\!/ H/\partial\!\!\!/\lambda)$.
Here, ${\bf z}_t$ denotes a trajectory evolving under Eq. (\ref{eq:modifiedEOM}) as $\lambda$ is varied from $A$ to $B$;
the integrand $\dot\lambda\,\partial\!\!\!/ H/\partial\!\!\!/\lambda$ is evaluated along this trajectory;
and $\langle\cdots\rangle_{\bf u}$ indicates an average over an ensemble of such trajectories, with initial conditions sampled from equilibrium.
Setting $t=\tau$ and integrating Eq. (\ref{FK}) over phase space, we obtain
\begin{equation}
\label{modifiedequation}
e^{-\beta\Delta F} = \left\langle e^{-\beta W} \right\rangle_{\bf u} \quad,
\end{equation}
where
\begin{equation}
\label{eq:wprime}
W = \int_0^\tau \dot{\lambda}\,
\frac{\partial\!\!\!/ H_{\lambda}}{\partial\!\!\!/\lambda}({\bf z}_t,\lambda_t) \, {\rm d}t
\end{equation}
is interpreted as the {\it work} performed on a system evolving under Eq. (\ref{eq:modifiedEOM}).

[We have derived Eq. (\ref{modifiedequation}) by equating two solutions of the sink equation (Eq.~\ref{sink}):
one obtained by inspection (Eq.~\ref{Boltzmann}), the other via path integration (right side of Eq.~\ref{FK}).
An alternative derivation proceeds by first {\it defining}
$g({\bf z},t) = \langle \delta({\bf z}-{\bf z}_t)\exp(-\beta w_t)\rangle_{\bf u}$,
then showing that this function satisfies Eq. (\ref{sink}), whose solution is in turn given by Eq. (\ref{Boltzmann}).
See Refs.~\cite{CJ:MasterEquation,Imparato05a} for analogous derivations of Eq. (\ref{previousresult}).]

Eq. (\ref{modifiedequation}) implies we can estimate $\Delta F$ by taking the exponential average of $W$ (Eq.~\ref{eq:wprime}), over trajectories evolving under the modified dynamics (Eq.~\ref{eq:modifiedEOM}).
This generalizes the usual fast switching method: we recover Eq.~\ref{previousresult} by choosing ${\bf u}={\bf 0}$.
Our approach also contains elements of both the {\it metric scaling}~\cite{Miller00} and {\it targeted perturbation}~\cite{CJ:Targetting,Biasedsampling} strategies, reducing to a variant of the former in the case of linear flow fields, ${\bf u} = \alpha(\lambda)\,{\bf z}$, and to the latter in the limit of instantaneous switching, $\tau\rightarrow 0$.
In that limit, the term $\tilde{\bf v}$ in Eq. (\ref{eq:modifiedEOM}) becomes negligible, and the trajectory evolves by integration along the flow field:
${\rm d}{\bf z}_\lambda/{\rm d}\lambda = {\bf u}({\bf z}_\lambda,\lambda)$.

While Eq. (\ref{modifiedequation}) is valid for any well-behaved flow field, the efficiency of the free energy estimate (the convergence of the exponential average) depends critically on the choice of ${\bf u}({\bf z},\lambda)$.
The challenge then is to construct flow fields that render calculations using Eq. (\ref{modifiedequation}) more efficient than those using Eq. (\ref{previousresult}).
Since the typically poor convergence of Eq. (\ref{previousresult}) correlates with the lag that develops between the current state of the system ($f$) and the instantaneous equilibrium distribution ($p^{\rm eq}$), {\it it is reasonable to speculate that a flow field ${\bf u}$ which reduces this lag will improve the efficiency of the free energy estimate}.
To pursue this idea, let us imagine for a moment that we are able to construct a {\it perfect} flow field, ${\bf u}^*$, that eliminates the lag entirely.
In this case the distribution $f({\bf z},t)=p^{\rm eq}({\bf z},\lambda_t)$ is a solution of Eq. (\ref{Fevolution}).
Substituting this solution into Eq. (\ref{Fevolution}), we get, using ${\cal L}_\lambda \cdot p^{\rm eq}=0$,
\begin{equation}
\label{eq:exactu0}
\frac{\partial p^{\rm eq}}{\partial\lambda} + \nabla\cdot\left({\bf u}^* p^{\rm eq}\right) = 0.
\end{equation}
Setting $p^{\rm eq}=e^{\beta(F-H)}$, we obtain
\begin{equation}
\label{exactu}
\frac{{\rm d}F}{{\rm d}\lambda}(\lambda) = 
\frac{\partial H}{\partial\lambda} + {\bf u}^* \cdot \nabla H - \beta^{-1} \nabla\cdot{\bf u}^* 
\equiv \frac{\partial\!\!\!/ H}{\partial\!\!\!/\lambda} ,
\end{equation}
therefore
\begin{equation}
\label{eq:Wstar}
W = \int_0^\tau \dot\lambda \frac{\partial\!\!\!/ H}{\partial\!\!\!/\lambda} \, {\rm d}t =
\int_0^\tau \dot\lambda \frac{{\rm d}F}{{\rm d}\lambda} \, {\rm d}t = \Delta F
\end{equation}
for every trajectory ${\bf z}_t$.
Thus, for a perfect flow field ${\bf u}^*$, there is no dissipation ($W_{\rm diss}=0$) and
{\it a single trajectory provides the correct free energy difference}.

Although on general grounds we expect that a perfect flow field typically exists,\footnote{
Since Eq. (\ref{eq:exactu0}) is of the form
$\nabla\cdot{\bf A} = q({\bf z},\lambda)$,
a formal solution can be constructed using Green's functions.
}
it seems unlikely we will be able to solve for ${\bf u}^*$ analytically, apart from a few simple systems.
(Indeed, Eq (\ref{exactu}) suggests that an expression for ${\rm d}F/{\rm d}\lambda$ is required to obtain ${\bf u}^*$.)
However, by revealing that {\it elimination} of the lag results in a zero-variance estimator of $\Delta F$, Eq. (\ref{eq:Wstar}) supports our earlier speculation: if we can construct a flow field that {\it reduces} the lag, 
then we should expect improved convergence of the exponential average.
We now illustrate this idea.
 
\begin{figure}[b]
\includegraphics[scale=0.3, angle=-90]{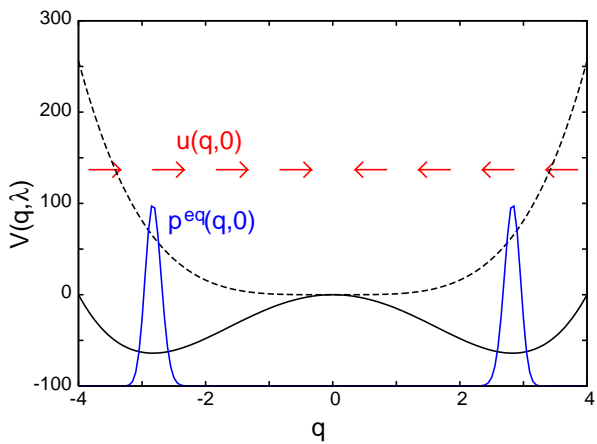}
\centering
\caption{The potential energy landscape for $\lambda=0$ (solid line) and $\lambda=1$ (dashed line).  Also depicted are the equilibrium distribution and the flow field, at $\lambda=0$.}
\label{fig:rev:potentialenergy}
\end{figure} 

Consider Sun's one-dimensional model system~\cite{SeanSun},
\begin{equation}
\label{quartichamiltonian}
H(p,q,\lambda)=\frac{p^2}{2m}+q^4-16(1- \lambda)q^2=\frac{p^2}{2m}+V(q,\lambda) .
\end {equation}
For $A\equiv 0\le\lambda<1\equiv B$, the potential energy profile $V(q,\lambda)$ is a double well, with minima at $\pm q_0(\lambda) \equiv  \pm \sqrt{8(1-\lambda)}$ separated by a barrier of height $64 (1-\lambda)^2$ (Fig.~\ref{fig:rev:potentialenergy}).
Setting $\beta=1$, the equilibrium distribution is bimodal and sharply peaked around $\pm q_0$; as $\lambda\rightarrow 1$ the two peaks coalesce as $V$ becomes a single, quartic well.
Analytical evaluation of the partition functions gives
$\Delta F = F_B - F_A = 62.94...$~\cite{Biasedsampling}.

\begin{figure}[tbp]
\includegraphics[scale=0.3, angle=-90]{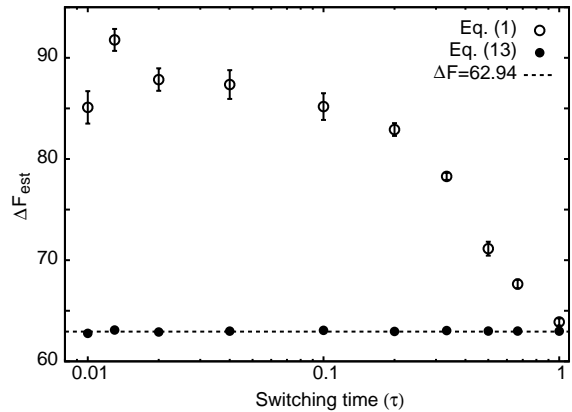}
\centering
\caption{
Comparison of estimates of $\Delta F$ using Eqs. (\ref{previousresult}) and (\ref{modifiedequation}).
We performed simulations for switching times ranging from $\tau=0.01$ to $\tau=1.0$.
Each $\Delta F_{\rm est}$ was obtained using $10^6$ trajectories, evolving under either Eq. (\ref{eq:physicalEOM}) (open circles) or Eqs. (\ref{eq:modifiedEOM}), (\ref{eq:tanhu}) (filled circles).
Error bars were computed using the bootstrap method; for the filled circles these were smaller than the symbols, and are not shown.
}
\label{fig:rev:freeenergyestimate}
\end{figure}

The direct application of Eq.~\ref{previousresult} to this model gives poor results when the switching is performed rapidly~\cite{SeanSun,Biasedsampling}.
A typical simulation begins with the system near $\pm q_0(0)$; then, as $\lambda$ is varied from 0 to 1, the two minima at $\pm q_0(\lambda)$ approach one another, but the system lags behind, resulting in large dissipation and poor free energy estimates.
This is illustrated by the open circles in Fig.~\ref{fig:rev:freeenergyestimate},
obtained from simulations during which the system evolved under Hamilton's equations, integrated using the velocity Verlet algorithm.
Only for $\tau =1$ does Eq. (\ref{previousresult}) provide an accurate estimate of $\Delta F$.
(The {\it systematic} error evident in Fig.~\ref{fig:rev:freeenergyestimate} arises after taking the logarithm of both sides of Eq. (\ref{previousresult})~\cite{Zuckerman02}.)

To illustrate the application of Eq. (\ref{modifiedequation}), 
let us take
\begin{equation}
\label{eq:tanhu}
u(q,\lambda) = \frac{{\rm d} q_0}{{\rm d} \lambda} \tanh \left[64 (1-\lambda) q_0 q\right] ,
\end{equation}
with $q_0=q_0(\lambda)$ as given above.
This field acts only on the coordinate $q$, and not on the momentum $p$.
We arrived at Eq. (\ref{eq:tanhu}) by using crude approximations to estimate the solution of Eq. (\ref{exactu}), modeling $p^{\rm eq}$ as a pair of Gaussians.
Omitting the details of this calculation, we note that near either peak of $p^{\rm eq}$, $u(q,\lambda)$ displaces the system toward the origin at a speed $\dot\lambda\vert u\vert \approx \dot\lambda \, {\rm d} q_0/{\rm d}\lambda$ (see Fig.~\ref{fig:rev:potentialenergy}).
This is the speed at which the two minima of $V(q,\lambda)$ approach the origin.
Intuitively, we expect this flow to reduce the lag between $f$ and $p^{\rm eq}$.

We repeated the simulations described above, now adding the term $\dot\lambda\,u$ to the dynamics.
The resulting estimates of $\Delta F$, obtained using Eq. (\ref{modifiedequation}) and depicted as filled circles in Fig.~\ref{fig:rev:freeenergyestimate}, are remarkably accurate over the entire range of switching times.
Indeed, for all $\tau = 0.01, \cdots, 1.0$, the work values $W$ were sharply peaked around $\Delta F$ (data not shown), confirming that the flow field escorts the system through a sequence of near-equilibrium states, even when $\lambda$ is switched rapidly.
We stress, however, that this choice of flow field is neither perfect ($u\ne u^*$) nor unique.
In particular, we expect it could be improved near $\lambda=1$, where the approximations made on the way to Eq. (\ref{eq:tanhu}) break down.

As illustrated by this simple, proof-of-principle example, the key to success with our method is a flow field ${\bf u}$ that reduces lag, and therefore dissipation, by mimicking the effect of a variation of $\lambda$ on the distribution $p^{\rm eq}$.
In general, such flow fields might be constructed using physical insight, experience and prior knowledge of the system, perhaps with iterative adjustment to improve convergence.~\footnote{
In this context, $\langle W\rangle$ represents a figure of merit~\cite{Miller00}: the smaller the average work, the better the flow field.}
We expect that, with trial and error, flow fields appropriate to a variety of problems will be developed.
For instance, (1) particle insertion into a fluid, (2) cavity growth in a fluid, and (3) the charging of a solute, all represent {\it in silico} thermodynamic processes for which we have some intuition regarding the atomic rearrangements that accompany the process.
Preliminary calculations, reported elsewhere~\cite{discreteflow}, confirm that such intuition can be used to design flow fields that significantly speed up the estimation of $\Delta F$.
Our method might also be combined with {\it steered molecular dynamics}~\cite{Izrailev97,Park04}, in which a constraining potential is used to drag a coordinate $\xi$ along a desired path $\tilde\xi_t$.
By adding a flow field that acts on this coordinate and others coupled to it, one might be able to reduce the lag between $\xi$ and $\tilde\xi_t$.
For free energy calculations along a reaction path for which we do not have good intuition, {\it transition path sampling}~\cite{TPS} could provide information useful for designing an effective flow field.

The method we propose is distinct from path-space sampling schemes~\cite{SeanSun,Geissler04,Wu05,Zuckerman}, in which the convergence of Eq. (\ref{previousresult}) is improved by modifying the {\it probabilities} with which physical trajectories are generated, for instance by biasing in favor of small work values.
In our approach, by contrast, we modify the equations of motion themselves, thereby sampling from an entirely different set of trajectories.
(Thus in the above example, we generated non-Hamiltonian trajectories, rather than a biased sampling of Hamiltonian trajectories.)
The distinction is particularly evident in the case of a perfect flow field ${\bf u}^*$, when {\it every} trajectory gives $W=\Delta F$.

Finally, while Eq. (\ref{modifiedequation}) is specifically a generalization of Eq. (\ref{previousresult}), the approach we take is readily extended to other nonequilibrium identities for free energy differences, including Crooks's fluctuation theorem~\cite{Crooks99} and Hummer and Szabo's identity for potentials of mean force~\cite{Hummer}.
Moreover, it would be interesting to combine our approach with the {\it large time step}~\cite{Largetimestep1} and {\it optimal protocol}~\cite{Schmiedl07} strategies, recently proposed for improving the efficiency of free energy estimates.

We gratefully acknowledge useful discussions with Andrew Ballard and Jordan Horowitz, and financial support provided by the University of Maryland.


\begin{thebibliography}{30}
\expandafter\ifx\csname natexlab\endcsname\relax\def\natexlab#1{#1}\fi
\expandafter\ifx\csname bibnamefont\endcsname\relax
  \def\bibnamefont#1{#1}\fi
\expandafter\ifx\csname bibfnamefont\endcsname\relax
  \def\bibfnamefont#1{#1}\fi
\expandafter\ifx\csname citenamefont\endcsname\relax
  \def\citenamefont#1{#1}\fi
\expandafter\ifx\csname url\endcsname\relax
  \def\url#1{\texttt{#1}}\fi
\expandafter\ifx\csname urlprefix\endcsname\relax\def\urlprefix{URL }\fi
\providecommand{\bibinfo}[2]{#2}
\providecommand{\eprint}[2][]{\url{#2}}

\bibitem[{\citenamefont{Frenkel and Smit}(2002)}]{Frenkel}
\bibinfo{author}{\bibfnamefont{D.}~\bibnamefont{Frenkel}} \bibnamefont{and}
  \bibinfo{author}{\bibfnamefont{B.}~\bibnamefont{Smit}},
  \emph{\bibinfo{title}{Understanding Molecular Simulation}}
  (\bibinfo{publisher}{Academic Press}, \bibinfo{address}{San Diego},
  \bibinfo{year}{2002}), \bibinfo{edition}{2nd} ed.

\bibitem[{\citenamefont{Chipot and Pohorille}(2007)}]{Chipot2007}
\bibinfo{author}{\bibfnamefont{C.}~\bibnamefont{Chipot}} \bibnamefont{and}
  \bibinfo{author}{\bibfnamefont{A.}~\bibnamefont{Pohorille}},
  \emph{\bibinfo{title}{Free Energy Calculations}}
  (\bibinfo{publisher}{Springer, Berlin}, \bibinfo{year}{2007}).

\bibitem[{\citenamefont{Jarzynski}(1997{\natexlab{a}})}]{CJ:Equality}
\bibinfo{author}{\bibfnamefont{C.}~\bibnamefont{Jarzynski}},
  \bibinfo{journal}{Phys. Rev. Lett.} \textbf{\bibinfo{volume}{78}},
  \bibinfo{pages}{2690} (\bibinfo{year}{1997}{\natexlab{a}}).

\bibitem[{\citenamefont{Hendrix and Jarzynski}(2001)}]{Hendrix2001}
\bibinfo{author}{\bibfnamefont{D.}~\bibnamefont{Hendrix}} \bibnamefont{and}
  \bibinfo{author}{\bibfnamefont{C.}~\bibnamefont{Jarzynski}},
  \bibinfo{journal}{J. Chem. Phys.} \textbf{\bibinfo{volume}{114}},
  \bibinfo{pages}{5974} (\bibinfo{year}{2001}).

\bibitem[{\citenamefont{Kofke}(2006)}]{Kofke}
\bibinfo{author}{\bibfnamefont{D.~A.} \bibnamefont{Kofke}},
  \bibinfo{journal}{Mol. Phys.} \textbf{\bibinfo{volume}{104}},
  \bibinfo{pages}{3701} (\bibinfo{year}{2006}), \bibinfo{note}{and references
  therein}.

\bibitem[{\citenamefont{Gore et~al.}(2003)\citenamefont{Gore, Ritort, and
  Bustamante}}]{Gore}
\bibinfo{author}{\bibfnamefont{J.}~\bibnamefont{Gore}},
  \bibinfo{author}{\bibfnamefont{F.}~\bibnamefont{Ritort}}, \bibnamefont{and}
  \bibinfo{author}{\bibfnamefont{C.}~\bibnamefont{Bustamante}},
  \bibinfo{journal}{Proc. Natl. Acad. Sci. U.S.A}
  \textbf{\bibinfo{volume}{100}}, \bibinfo{pages}{12564}
  (\bibinfo{year}{2003}).

\bibitem[{\citenamefont{Jarzynski}(2006)}]{CJ:Rareevents}
\bibinfo{author}{\bibfnamefont{C.}~\bibnamefont{Jarzynski}},
  \bibinfo{journal}{Phys. Rev. E} \textbf{\bibinfo{volume}{73}},
  \bibinfo{pages}{046105} (\bibinfo{year}{2006}).

\bibitem[{\citenamefont{Pearlman and Kollman}(1989)}]{Lag1}
\bibinfo{author}{\bibfnamefont{D.}~\bibnamefont{Pearlman}} \bibnamefont{and}
  \bibinfo{author}{\bibfnamefont{P.}~\bibnamefont{Kollman}},
  \bibinfo{journal}{J. Chem. Phys} \textbf{\bibinfo{volume}{91}},
  \bibinfo{pages}{7831} (\bibinfo{year}{1989}).

\bibitem[{\citenamefont{Wood}(1991)}]{Lag2}
\bibinfo{author}{\bibfnamefont{R.}~\bibnamefont{Wood}}, \bibinfo{journal}{J.
  Phys. Chem} \textbf{\bibinfo{volume}{95}}, \bibinfo{pages}{4838}
  (\bibinfo{year}{1991}).

\bibitem[{\citenamefont{Hermans}(1991)}]{Hermans91}
\bibinfo{author}{\bibfnamefont{J.}~\bibnamefont{Hermans}}, \bibinfo{journal}{J.
  Phys. Chem.} \textbf{\bibinfo{volume}{95}}, \bibinfo{pages}{9029}
  (\bibinfo{year}{1991}).

\bibitem[{\citenamefont{Vaikuntanathan and Jarzynski}()}]{discreteflow}
\bibinfo{author}{\bibfnamefont{S.}~\bibnamefont{Vaikuntanathan}}
  \bibnamefont{and}
  \bibinfo{author}{\bibfnamefont{C.}~\bibnamefont{Jarzynski}},
  \bibinfo{note}{unpublished}.

\bibitem[{\citenamefont{Jarzynski}(1997{\natexlab{b}})}]{CJ:MasterEquation}
\bibinfo{author}{\bibfnamefont{C.}~\bibnamefont{Jarzynski}},
  \bibinfo{journal}{Phys. Rev. E} \textbf{\bibinfo{volume}{56}},
  \bibinfo{pages}{5018} (\bibinfo{year}{1997}{\natexlab{b}}).

\bibitem[{\citenamefont{Hummer and Szabo}(2001)}]{Hummer}
\bibinfo{author}{\bibfnamefont{G.}~\bibnamefont{Hummer}} \bibnamefont{and}
  \bibinfo{author}{\bibfnamefont{A.}~\bibnamefont{Szabo}},
  \bibinfo{journal}{Proc. Natl. Acad. Sci. U.S.A}
  \textbf{\bibinfo{volume}{98}}, \bibinfo{pages}{3658} (\bibinfo{year}{2001}).

\bibitem[{\citenamefont{Hummer and Szabo}(2005)}]{Hummer05}
\bibinfo{author}{\bibfnamefont{G.}~\bibnamefont{Hummer}} \bibnamefont{and}
  \bibinfo{author}{\bibfnamefont{A.}~\bibnamefont{Szabo}},
  \bibinfo{journal}{Acc. Chem. Res.} \textbf{\bibinfo{volume}{38}},
  \bibinfo{pages}{504 } (\bibinfo{year}{2005}).

\bibitem[{\citenamefont{Ge and Jiang}(2008)}]{Ge08}
\bibinfo{author}{\bibfnamefont{H.}~\bibnamefont{Ge}} \bibnamefont{and}
  \bibinfo{author}{\bibfnamefont{D.-Q.} \bibnamefont{Jiang}},
  \bibinfo{journal}{J. Stat. Phys.} \textbf{\bibinfo{volume}{131}},
  \bibinfo{pages}{675 } (\bibinfo{year}{2008}).

\bibitem[{\citenamefont{Imparato and Peliti}(2005)}]{Imparato05a}
\bibinfo{author}{\bibfnamefont{A.}~\bibnamefont{Imparato}} \bibnamefont{and}
  \bibinfo{author}{\bibfnamefont{L.}~\bibnamefont{Peliti}},
  \bibinfo{journal}{Phys. Rev. E} \textbf{\bibinfo{volume}{72}},
  \bibinfo{pages}{046114} (\bibinfo{year}{2005}).

\bibitem[{\citenamefont{Miller and Reinhardt}(2000)}]{Miller00}
\bibinfo{author}{\bibfnamefont{M.~A.} \bibnamefont{Miller}} \bibnamefont{and}
  \bibinfo{author}{\bibfnamefont{W.~P.} \bibnamefont{Reinhardt}},
  \bibinfo{journal}{J. Chem. Phys.} \textbf{\bibinfo{volume}{113}},
  \bibinfo{pages}{7035 } (\bibinfo{year}{2000}).

\bibitem[{\citenamefont{Jarzynski}(2002)}]{CJ:Targetting}
\bibinfo{author}{\bibfnamefont{C.}~\bibnamefont{Jarzynski}},
  \bibinfo{journal}{Phys. Rev. E} \textbf{\bibinfo{volume}{65}},
  \bibinfo{pages}{046122} (\bibinfo{year}{2002}).

\bibitem[{\citenamefont{Oberhofer et~al.}(2005)\citenamefont{Oberhofer,
  Dellago, and Geissler}}]{Biasedsampling}
\bibinfo{author}{\bibfnamefont{H.}~\bibnamefont{Oberhofer}},
  \bibinfo{author}{\bibfnamefont{C.}~\bibnamefont{Dellago}}, \bibnamefont{and}
  \bibinfo{author}{\bibfnamefont{P.}~\bibnamefont{Geissler}},
  \bibinfo{journal}{J. Phys. Chem B} \textbf{\bibinfo{volume}{109}},
  \bibinfo{pages}{6902} (\bibinfo{year}{2005}).

\bibitem[{\citenamefont{Sun}(2003)}]{SeanSun}
\bibinfo{author}{\bibfnamefont{S.~X.} \bibnamefont{Sun}}, \bibinfo{journal}{J.
  Chem. Phys} \textbf{\bibinfo{volume}{118}}, \bibinfo{pages}{5759}
  (\bibinfo{year}{2003}).

\bibitem[{\citenamefont{Zuckerman and Woolf}(2002)}]{Zuckerman02}
\bibinfo{author}{\bibfnamefont{D.~M.} \bibnamefont{Zuckerman}}
  \bibnamefont{and} \bibinfo{author}{\bibfnamefont{T.~B.} \bibnamefont{Woolf}},
  \bibinfo{journal}{Phys. Rev. Lett.} \textbf{\bibinfo{volume}{89}},
  \bibinfo{pages}{180602} (\bibinfo{year}{2002}).

\bibitem[{\citenamefont{Izrailev et~al.}(1997)\citenamefont{Izrailev,
  Stepaniants, Balsera, Oono, and Schulten}}]{Izrailev97}
\bibinfo{author}{\bibfnamefont{S.}~\bibnamefont{Izrailev}},
  \bibinfo{author}{\bibfnamefont{S.}~\bibnamefont{Stepaniants}},
  \bibinfo{author}{\bibfnamefont{M.}~\bibnamefont{Balsera}},
  \bibinfo{author}{\bibfnamefont{Y.}~\bibnamefont{Oono}}, \bibnamefont{and}
  \bibinfo{author}{\bibfnamefont{K.}~\bibnamefont{Schulten}},
  \bibinfo{journal}{Biophys. J} \textbf{\bibinfo{volume}{72}},
  \bibinfo{pages}{1568 } (\bibinfo{year}{1997}).

\bibitem[{\citenamefont{Park and Schulten}(2004)}]{Park04}
\bibinfo{author}{\bibfnamefont{S.}~\bibnamefont{Park}} \bibnamefont{and}
  \bibinfo{author}{\bibfnamefont{K.}~\bibnamefont{Schulten}},
  \bibinfo{journal}{J. Chem. Phys.} \textbf{\bibinfo{volume}{120}},
  \bibinfo{pages}{5946 } (\bibinfo{year}{2004}).

\bibitem[{\citenamefont{Dellago et~al.}(1998)\citenamefont{Dellago, Bolhuis,
  Csajka, and Chandler}}]{TPS}
\bibinfo{author}{\bibfnamefont{C.}~\bibnamefont{Dellago}},
  \bibinfo{author}{\bibfnamefont{P.~G.} \bibnamefont{Bolhuis}},
  \bibinfo{author}{\bibfnamefont{F.~S.} \bibnamefont{Csajka}},
  \bibnamefont{and} \bibinfo{author}{\bibfnamefont{D.}~\bibnamefont{Chandler}},
  \bibinfo{journal}{J. Chem. Phys.} \textbf{\bibinfo{volume}{108}},
  \bibinfo{pages}{1964 } (\bibinfo{year}{1998}).

\bibitem[{\citenamefont{Geissler and Dellago}(2004)}]{Geissler04}
\bibinfo{author}{\bibfnamefont{P.~L.} \bibnamefont{Geissler}} \bibnamefont{and}
  \bibinfo{author}{\bibfnamefont{C.}~\bibnamefont{Dellago}},
  \bibinfo{journal}{J. Phys. Chem. B} \textbf{\bibinfo{volume}{108}},
  \bibinfo{pages}{6667 } (\bibinfo{year}{2004}).

\bibitem[{\citenamefont{Wu and Kofke}(2005)}]{Wu05}
\bibinfo{author}{\bibfnamefont{D.}~\bibnamefont{Wu}} \bibnamefont{and}
  \bibinfo{author}{\bibfnamefont{D.~A.} \bibnamefont{Kofke}},
  \bibinfo{journal}{J. Chem. Phys.} \textbf{\bibinfo{volume}{122}},
  \bibinfo{pages}{204104} (\bibinfo{year}{2005}).

\bibitem[{\citenamefont{Ytreberg and Zuckerman}(2004)}]{Zuckerman}
\bibinfo{author}{\bibfnamefont{F.~M.} \bibnamefont{Ytreberg}} \bibnamefont{and}
  \bibinfo{author}{\bibfnamefont{D.~M.} \bibnamefont{Zuckerman}},
  \bibinfo{journal}{J. Chem. Phys} \textbf{\bibinfo{volume}{120}},
  \bibinfo{pages}{10876} (\bibinfo{year}{2004}).

\bibitem[{\citenamefont{Crooks}(1999)}]{Crooks99}
\bibinfo{author}{\bibfnamefont{G.~E.} \bibnamefont{Crooks}},
  \bibinfo{journal}{Phys. Rev. E} \textbf{\bibinfo{volume}{60}},
  \bibinfo{pages}{2721 } (\bibinfo{year}{1999}).

\bibitem[{\citenamefont{Lechner et~al.}(2006)\citenamefont{Lechner, Oberhofer,
  Dellago, and Geissler}}]{Largetimestep1}
\bibinfo{author}{\bibfnamefont{W.}~\bibnamefont{Lechner}},
  \bibinfo{author}{\bibfnamefont{H.}~\bibnamefont{Oberhofer}},
  \bibinfo{author}{\bibfnamefont{C.}~\bibnamefont{Dellago}}, \bibnamefont{and}
  \bibinfo{author}{\bibfnamefont{P.}~\bibnamefont{Geissler}},
  \bibinfo{journal}{J. Chem. Phys} \textbf{\bibinfo{volume}{124}},
  \bibinfo{pages}{044113} (\bibinfo{year}{2006}).

\bibitem[{\citenamefont{Schmiedl and Seifert}(2007)}]{Schmiedl07}
\bibinfo{author}{\bibfnamefont{T.}~\bibnamefont{Schmiedl}} \bibnamefont{and}
  \bibinfo{author}{\bibfnamefont{U.}~\bibnamefont{Seifert}},
  \bibinfo{journal}{Phys. Rev. Lett.} \textbf{\bibinfo{volume}{98}},
  \bibinfo{pages}{108301} (\bibinfo{year}{2007}).

\end{thebibliography}
\end{document}